\documentclass[journal,twocolumn]{IEEEtran}

\usepackage{balance}

\usepackage{amsmath,amssymb,amsfonts}
\usepackage{accents}
\usepackage{algorithmic}
\usepackage{graphicx}
\usepackage{textcomp}
\usepackage{mathrsfs}
\usepackage{mathtools}
\usepackage[T1]{fontenc}
\usepackage{bbm} 
\usepackage{bbding}
\usepackage{keystroke}
\usepackage{pifont}
\usepackage{relsize}
\usepackage{amsthm}
\usepackage{amsmath}
\usepackage{mathrsfs}
\usepackage{amssymb} 
\usepackage{stackrel}
\usepackage{mathtools}
\usepackage{nicefrac}
\usepackage{verbatim}
\usepackage{enumerate}

\usepackage[square,numbers,sort&compress]{natbib}
\usepackage{xcolor}
\usepackage{hyperref}
\hypersetup{
    colorlinks=true,
    linkcolor=blue,
    citecolor=blue,
    urlcolor=cyan
}

\newcommand{\Eq}[1]{Eq.~(\ref{#1})}
\newcommand{\Fig}[1]{Fig.~\ref{#1}}

\newcommand{\Sec}[1]{Sec.~\ref{#1}}
\newcommand{\cRef}[1]{Ref.\cc{#1}}
\newcommand{\cRefs}[1]{Refs.~\cc{#1}}
\newcommand{\cc}[1]{~\mbox{\cite{#1}}}

\newcommand{\Thm}[1]{Theorem~\ref{#1}}
\newcommand{\Prop}[1]{Proposition~\ref{#1}}
\newcommand{\App}[1]{Appendix~\ref{#1}}

\theoremstyle{remark}\newtheorem{theorem}{Theorem}
\theoremstyle{remark}
\theoremstyle{remark}
\theoremstyle{remark}\newtheorem{proposition}[theorem]{Proposition}
\theoremstyle{remark}
\theoremstyle{remark}\newtheorem{remark}{Remark}
\theoremstyle{remark}

\newcommand{\norm}[1]{{\| #1 \|}}  
\newcommand{\ket}[1]{{ |{#1} \rangle }}  
\newcommand{\bra}[1]{{ \langle {#1} | }}

\newcommand{\ketbra}[2]{{ |{#1} \rangle\!\,\langle {#2} | }}
\newcommand{\fqswEpsilon}[1]{\mathrel{\stackrel{\mathclap{\textstyle #1}}{\approx}}}
\newcommand{\ubar}[1]{\underaccent{\bar}{#1}}

\newcommand{\plainfootnote}[1]{%
  \renewcommand{\thefootnote}{}%
  \footnote{#1}%
  \addtocounter{footnote}{-1}%
}

\newcommand{\EqDef}{\stackrel{\mathrm{def}}{=}}
\newcommand{\Id}{\mathbbm{1}}

\DeclareMathOperator{\Tr}{Tr}

\begin{document}

\title{Quantum Information Transmission Over\\ Quantum Relay Channels}

\author{Yigal Ilin and Uzi Pereg\\
\IEEEauthorblockA{\textit{ECE Department \& Helen Diller Quantum Center,
Technion}\\
\tt yigal.ilin@gmail.com, uzipereg@technion.ac.il}}

\maketitle

\begin{abstract}
Quantum relays are central to both quantum communication and distributed quantum computing, enabling long-distance transmission and modular architectures. Unlike classical repeaters, quantum repeaters preserve coherence without amplifying quantum information, relying on entanglement swapping and quantum error correction to overcome loss and decoherence. In this work, we investigate the transmission of quantum information via quantum relay channels. Our three-terminal relay model captures the trade-off between repeater-assisted and repeaterless communication strategies. Specifically, we propose a partial decode-forward strategy, in which quantum ``message system" consists of two components. The first component is decoded by the relay and then sent to the destination receiver, whereas the second component is decoded by the destination receiver without the relay's help. We analyze both entanglement-assisted and unassisted scenarios. As a special case, the full decode-forward strategy is recovered, with the relay decoding, re-encoding, and forwarding the entire message. Our framework allows for different entanglement topologies between the transmitter, the relay and the destination receiver, recovering known results on entanglement-assisted and unassisted communication. Furthermore, we discuss the interpretation of coding with quantum side information. These findings provide a foundation for designing secure, efficient, and reliable quantum networks and for realizing practical quantum repeaters and long-range quantum key distribution.
\end{abstract}



\IEEEpeerreviewmaketitle

\section{Introduction}\label{sec:introduction}

\plainfootnote{Part of this work has been presented 
at the 61st Allerton Conference on Communication, Control, and Computing, Urbana, IL, September 2025.}

Relaying quantum information plays a central role in quantum communication and distributed quantum computing, enabling  long-distance transmission  and modular architectures\cc{ref:Bloch2014, ref:Hayashi2021}. Quantum information technology faces significant challenges due to the fragile nature of quantum states, which are highly susceptible to noise. In quantum computing, fault tolerance requires robust quantum error correction (QEC) codes and efficient syndrome extraction methods\cc{ref:GoogleThreshold2024}. However, practical challenges such as correlated errors and slow decoding limit current QEC performance\cc{ref:Chamberland2023-backlog}. Distributed relays could facilitate high-rate quantum error correction by enabling concurrent decoding across separate quantum devices\cc{ref:Skoric2023}, yet demand coherent and reliable quantum state transfer across them\cc{ref:Caleffi2024}. Moreover, reliable long-range quantum communication is severely limited by photon loss and environmental noise, in both fiber-optic and free-space transmission\cc{ref:Mele2024}. By leveraging entanglement swapping\cc{ref:Ghaderibaneh2022} and teleportation-based protocols\cc{ref:Cacciapuoti2020}, quantum relays not only help maintain coherence and entanglement in distributed architectures, but also extend the distance between end users\cc{ref:Vu2020, ref:Liorni2021, ref:Baskar2024}. 

\begin{figure}[b]
  \centering{\includegraphics[width=1\linewidth]{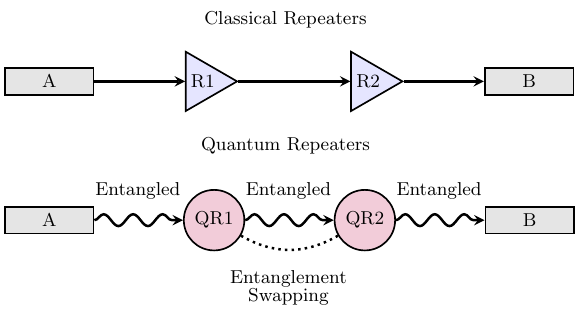}}
   \caption{Comparison of classical and quantum repeater architectures. Classical repeaters amplify signals, while quantum repeaters utilize entanglement.}
    \label{fig:repeater-intuition}
\end{figure}

From a security standpoint, a the US National Security Agency (NSA) has highlighted several challenges regarding the current applicability of quantum cryptography\cc{ref:Nsa-White-Paper}. Among these challenges is the need for \emph{trusted} intermediate stations in current implementations of quantum key distribution (QKD). As pointed out by Renner and Wolf  \cc{ref:Renner-Wolf-Nsa-Resp, ref:Renner2023}, this will change once quantum repeaters are employed\cc{ref:Briegel1998}. In other words, quantum repeaters enable the extension of communication distances without requiring the intermediate relay nodes to be trusted. These devices work entirely on the quantum level and hence do not need to be trusted. Therefore, using quantum repeaters will prevent security risks from insider threats. Understanding the relaying of quantum information over quantum channels can serve as a significant milestone on the road to developing quantum repeaters.
One practical approach to extending the achievable distance  is twin-field QKD, which improves the secret key rate scaling from $O(\sqrt\eta)$ to $O(\eta)$, where $\eta$ is the channel transmissivity \cite{maeda2019repeaterless}. However, since the transmissivity itself decreases exponentially with distance, the achievable range remains fundamentally limited. Therefore, quantum repeaters are still indispensable for long-distance quantum communication.

In order to distribute entanglement and send quantum information, the quantum relay would need to operate as a quantum repeater\cc{ref:Zoller1998}. A quantum repeater is a device that enables long-distance quantum communication by dividing the transmission into shorter segments and using entanglement swapping and purification to extend quantum correlations without directly amplifying quantum information, which is fundamentally prohibited by the no-cloning theorem. Unlike classical repeaters, they do not amplify signals but instead rely on QEC or entanglement-based techniques to overcome loss and decoherence\cc{ref:Azuma2015, ref:Azuma2023}. A  basic comparison between the classical and quantum repeaters is illustrated in \Fig{fig:repeater-intuition}.

Cooperation in quantum communication networks has become a major focus of research in recent years, driven by advances in both experimental techniques and theoretical developments\cc{ref:vanLoockAltBecherBensonBocheDeppe:20p,ref:luo2023recent,ref:nemirovsky2024increasing}. 
Entanglement serves as a valuable resource for network communication\cc{ref:BFSDBFJ:20b}. In the point-to-point setting, entanglement assistance between the transmitter and receiver can significantly boost throughput\cc{ref:BennettShorSmolin:02p}, even when the entanglement resource is noisy\cc{ref:ZhuangZhuShor:17p} or unreliable\cc{ref:PeregDeppeBoche:23p}.

In general, the quantum communication literature can broadly be categorized into two  approaches: repeater-aided and repeaterless transmission. The Pirandola–Laurenza–Ottaviani–Banchi (PLOB) bound\cc{ref:Pirandola2017-PLOB}
sets the fundamental rate-distance limit  for repeaterless communication, i.e., without a quantum repeater. The repeater-aided transmission enables the distribution of quantum correlations across long distances, by dividing the channel into shorter segments and placing  where intermediate relay terminals. Through entanglement swapping, the quantum repeaters generate entanglement and maintain coherence without violating the no-cloning theorem\cc{ref:Pirandola2019, ref:Winnel2021}. Experimental realizations for repeaterless and repeater-aided communication were demonstrated in \cRefs{ref:Lucamarini2018} and \cc{ref:Pittaluga2025}, respectively.

\begin{figure}[t]
  \centering{\includegraphics[width=0.55\linewidth]{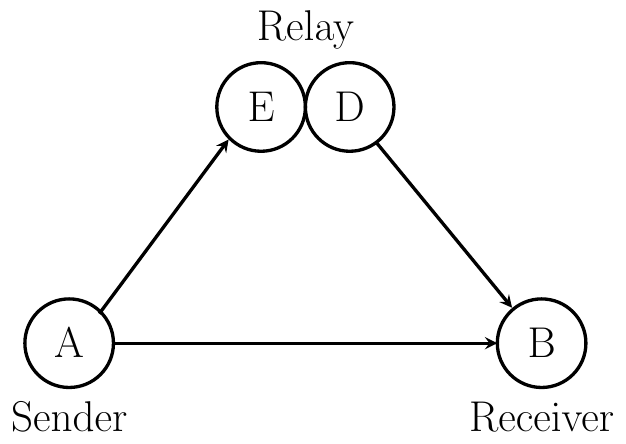}}
   \caption{Three-terminal quantum relay schematic. The transmitters at the sender and relay are denoted by $A$ and $D$, respectively, while the receivers at the relay and destination are labeled $E$ and $B$.}
    \label{fig:qrelay-three-party}
\end{figure}

\begin{figure}[t]
  \centering{\includegraphics[width=1\linewidth]{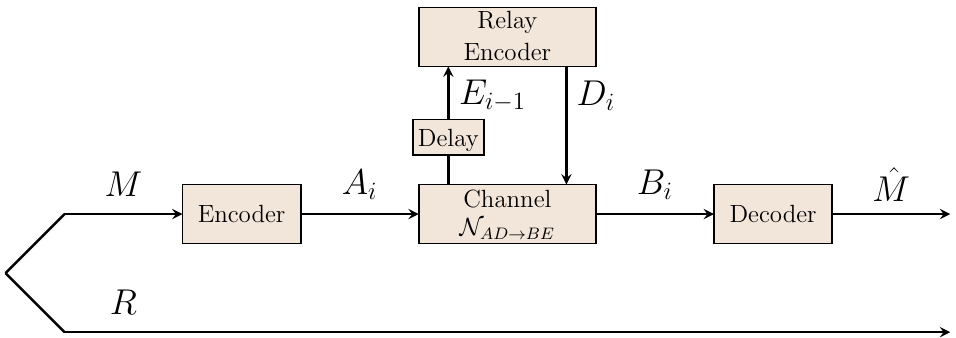}}
   \caption{Coding for a quantum relay channel $\mathcal{N}_{AD \rightarrow BE}$.}
    \label{fig:qrelay-comm-scheme}
\end{figure}

A schematic diagram of the quantum relay channel is shown in 
\Fig{fig:qrelay-three-party}.
In multihop networks, multiple rounds of communication take place, making synchronization a crucial requirement\cc{ref:KramerGastparGupta:05p,ref:Kramer:08n}. Consider a central clock that ticks $n$ times. Between clock ticks $i-1$ and $i$, the sender and relay transmit their respective channel inputs, $A_i$ and $D_i$ (see \Fig{fig:qrelay-comm-scheme}). At tick $i$, the relay and destination receive the channel outputs, $E_i$ and $B_i$, respectively. A short delay is thus necessary to ensure causal operation. The classical relay channel model was originally introduced by van der Meulen\cc{ref:vanderMeulen:71p}, and later adopted as a foundation for multihop network models\cc{ref:PeregSteinberg:19p3}.

Causality constraints in network settings often require the use of block Markov coding\cc{ref:ChoudhuriKimMitra:13p}, where the transmitter sends a sequence of blocks, and each block encodes information related to both current and previous transmissions. Quantum versions of block Markov coding have been studied in the context of classical-quantum (c-q) relay channels\cc{ref:SavovWildeVu:12c,ref:Savov:12z}, quantum communication with parameter estimation\cc{ref:Pereg:21p}, and quantum cribbing scenarios\cc{ref:PeregDeppeBoche:22p}. 
An information-theoretic framework for primitive quantum relay channels was first introduced in\cc{ref:Pereg2021-Prim-Relay}, and later extended in\cc{ref:Pereg2024-Relay} to address classical communication over fully quantum relay channels using various coding strategies. Other quantum repeater models can be found in\cc{ref:Ghalaii2020, ref:Gyongyosi2012, ref:Pirandola2016, ref:Gyongyosi2014, ref:Shi2012}.

In this work, we focus on the transmission of \emph{quantum} information over quantum relay channels, aiming to characterize the trade-offs between repeater-assisted and repeaterless communication scenarios. 
We study the three-terminal quantum relay channel illustrated in \Fig{fig:qrelay-three-party}. This type of relay network serves as a fundamental building block in cooperative communication \cc{ref:Chakrabarti2006} and can also be used to establish entanglement between network nodes \cc{ref:NatorPereg2024}.
While the previous work on quantum relay channels in \cc{ref:Pereg2024-Relay} addressed the transmission of classical messages, here we study quantum subspace transmission, i.e., the reliable transmission of quantum information. This transition fundamentally changes both the performance metric and the proof techniques. Instead of mutual information and Holevo information expressions established via packing lemma and typicality arguments, the achievable rates are characterized in terms of coherent information, and the analysis relies on the decoupling approach \cite{ref:Dupuis2009}.
Moreover, unlike classical communication, reliable quantum information transmission inherently provides secrecy. By the monogamy of entanglement, successful transmission implies that the environment cannot simultaneously retain significant information about the transmitted quantum state.

The main technical contribution of this work is a novel coding scheme that combines block Markov coding \cite{ref:Pereg2024-Relay} with the decoupling approach, which forms the basis of our analysis.
Specifically, we consider the partial decode-forward strategy, in which quantum ``message system" consists of two components.
The first component is decoded by the relay and then sent to the destination receiver, whereas the second message component is decoded by the destination receiver without the relay's help. We derive achievable rates, both with and without entanglement assistance. As a special case, we obtain an achievable rate corresponding to a full decode-forward strategy, in which the relay  decodes the entire information, re-encodes it, and forwards it to the destination receiver. In the unassisted case, we leverage the fact that communication occurs over multiple rounds, allowing the protocol to reuse entanglement generated as a by-product in block $i-1$ during block $i$.


We show that the quantum rate $Q$ below is achievable for the quantum relay channel:
\begin{multline}
 Q= \max_{\sigma} \bigl[ \min \bigl\{ I(A_1 A_2\rangle B)_\omega \,,\\\; I(A_1\rangle E)_\omega  + I(A_2\rangle B  A_1)_\omega \bigr\}\bigr]
\end{multline}
where $I(A\rangle B)_\rho$ denotes the coherent information corresponding to a quantum state $\rho_{AB}$.
Here, 
$A$, $D$, $E$, and $B$ represent the sender's transmitter, the relay's transmitter, the relay's receiver, and the destination receiver, respectively (see \Fig{fig:qrelay-three-party}). The maximum is taken over all $\ket{\sigma_{A_1 A_2 AD}}$, where $A_1$ and $A_2$ are ancillas that represent the quantum message components mentioned above.  Intuitively, $A_1$ represents Alice's share of entanglement with the relay, while $A_2$ is Alice's share of entanglement with Bob. Due to entanglement monogamy\cc{ref:Brando2017}, there is a tradeoff between the entanglement that Alice generates with the relay and with the destination decoder.

Our model also permits entanglement between the inputs of Alice and the relay (see \Fig{fig:qrelay-comm-scheme}), which as we show leads to improved achievable transmission rates over the case of the orthogonal links (see \cRef{ref:Ghalaii2020} for an example of a repeater chain composed of orthogonal links). The capacity bound with entanglement assistance resembles that  of Dupuis \cRef{ref:Dupuis2009}, who studied the entanglement-assisted capacity of quantum channels with side information at the transmitter.


While QEC has advanced significantly\cc{ref:Laflamme1996, ref:Knill1997, ref:Postler2024}, practical bottlenecks remain, especially the need for rapid, accurate decoding to avoid syndrome backlogs that can exponentially slow computation\cc{ref:Terhal2015, ref:Holmes2020-backlog}. A promising approach is to parallelize syndrome extraction across distributed devices, where ancillary registers are connected through quantum relays enabling reliable state transfer via entanglement swapping and teleportation\cc{ref:Goebel2008, ref:deRiedmatten2004, ref:Metcalf2014}. In that context, our setup can also be interpreted as follows. Consider the communication setup depicted in \Fig{fig:qrelay-comm-scheme}, the undisturbed reference system $R$ corresponds to the quantum data register, and the ancillary system $M$, transmitted through the repeater-assisted communication channel enables subsequent error syndrome extraction. This distributed approach could enhance the performance of fault-tolerant architectures, paving the way toward more scalable, higher-fidelity quantum error correction in large quantum systems\cc{ref:Battistel2023, ref:Wang2023}.

The paper is organized as follows. In \Sec{sec:preliminaries-notation}, we present preliminary definitions and the quantum relay channel model. \Sec{sec:coding-scheme} describes the coding framework for strictly causal quantum relaying. Our main results for the partial and full decode-forward strategies, including assisted and unassisted scenarios, are given in \Sec{sec:main-results}. Then, in \Sec{sec:anal-prop-1} we give an intuitive analysis of our main results. \Sec{sec:quant-relay-examples} provides an example based on the quantum erasure channel. \Sec{sec:summary-conclusion} provides a summary and concluding remarks. Finally, \App{sec:fqsw-thrm} overviews the fully quantum Slepian-Wolf (FQSW) protocol, central to our analysis, while detailed proofs of the achievable rates are given in \App{sec:proof-prop-1} and \App{sec:proof-theorems-2-3-4}.

\section{Preliminaries and Notation}\label{sec:preliminaries-notation}

In this section, we introduce the basic definitions and notation used throughout the paper. A quantum state is represented by a density operator $\rho$ acting on a Hilbert space $\mathcal{H}$, satisfying $\Tr(\rho) = 1$ and $\rho \succeq 0$. We assume all Hilbert spaces to be finite-dimensional. The set of all such density operators is denoted by $\mathcal{D}(\mathcal{H})$.

The outcome distribution of a quantum measurement can be described using a positive operator-valued measure (POVM), which is a collection of positive semi-definite operators $\{\Delta_j\}$ such that $\sum_j \Delta_j = \Id$, where $\Id$ is the identity operator. According to the Born rule, if the system is in the state $\rho$, the probability of observing outcome $j$ is given by
$%
    \Pr(j) = \Tr(\Delta_j \rho).
$ 
The quantum entropy of a density operator $\rho$ is defined as
$
    H(\rho) \EqDef -\Tr(\rho \log \rho),
$ 
which reduces to the Shannon entropy corresponding to the eigenvalue distribution of $\rho$. For a bipartite quantum state $\rho_{AB} \in \mathcal{D}(\mathcal{H}_A \otimes \mathcal{H}_B)$, the quantum mutual information is defined as
\begin{equation*}
    I(A;B)_\rho = H(\rho_A) + H(\rho_B) - H(\rho_{AB}),
\end{equation*}
where $\rho_A = \Tr_B(\rho_{AB})$ and $\rho_B = \Tr_A(\rho_{AB})$ are the reduced density operators. The conditional quantum entropy is defined as
$
    H(A|B)_\rho = H(\rho_{AB}) - H(\rho_B)
$, 
and the coherent information as
$
I(A\rangle B)_\rho 
= H(\rho_B)-H(\rho_{AB})_\rho
= -H(A|B)_\rho
$. 

We use the following notation for state approximation: We write
$    \rho\fqswEpsilon{\varepsilon}\sigma$ iff
$   \norm{\rho-\sigma}_1 \leq\varepsilon$.
 Uhlmann's theorem is a useful tool in this work: consider the pure states $\ket{\varphi}_{AB}$ and $\ket{\psi}_{AC}$, where $\ket{\psi}_{AC}$ is non-normalized. If the reduced states satisfy $\varphi_A\fqswEpsilon{\varepsilon}\psi_A$,
then 
there exists a partial isometry $V_{C \rightarrow B}$ such that
\begin{equation*}
    \ket{\varphi}_{AB} \fqswEpsilon{4\sqrt{\varepsilon}}
    V_{C \rightarrow B} \ket{\psi}_{AC}
    .
\end{equation*}
Note that this version of the Uhlmann's theorem is essentially a generalization of the original theorem to the case where the purifications are allowed to be non-normalized, introducing an additional factor of $2$ into the above approximation (see Lemma 5 in \cRef{ref:Berta2017}).

In this work, we consider a fully quantum relay channel as a model for the three-terminal network depicted in \Fig{fig:qrelay-three-party}. A quantum relay channel is described by a completely-positive trace-preserving (CPTP) map $\mathcal{N}_{AD \rightarrow BE}$, where $A$, $D$, $B$, and $E$ correspond to the sender's transmitter, the relay's transmitter, the destination receiver, and the relay's receiver, respectively (see \Fig{fig:qrelay-three-party}).
We assume that the channel is memoryless. Specifically, if the sender and relay inputs are given by the systems $A^n = (A_1, A_2, \ldots, A_n)$ and $D^n = (D_1, D_2, \ldots, D_n)$ across $n$ channel uses, then the joint input state $\rho_{A^n D^n}$ evolves under the tensor product channel
$
    \mathcal{N}_{A^n D^n \rightarrow B^n E^n} \EqDef \left( \mathcal{N}_{AD \rightarrow BE} \right)^{\otimes n}
$. 
The specific communication setting will be formalized in \Sec{sec:coding-scheme}, where we require that the relay encodes information in a \emph{strictly causal} manner. That is, at each time step $i$, the relay first transmits $D_i$ and only then receives $E_i$.

\section{Coding Scheme}\label{sec:coding-scheme}

Consider a quantum relay channel $\mathcal{N}_{AD\rightarrow BE}$. We define a code for entanglement generation 
via $\mathcal{N}_{AD\rightarrow BE}$, from Alice to Bob. A  $(2^{nQ},n)$ quantum code for 
$\mathcal{N}_{AD\rightarrow BE}$ is defined by:

\begin{itemize}
    \item A Hilbert space $\mathcal{H}_M$ of dimension $\dim(\mathcal{H}_M)=2^{nQ}$.

    \item An encoding channel $\mathcal{E}_{M\rightarrow A^{n}}:\mathcal{D}(\mathcal{H}_M)\rightarrow\mathcal{D}(\mathcal{H}_{A}^{\otimes n})$, which encodes the state of the information system $M$ into the input system $A^n$. 

    \item A sequence of strictly casual encoding maps $\Gamma^{(i)}_{E_{i-1}\bar{E}^{i-2}\rightarrow{\bar{E}^{i-1}D_{i}}}:\mathcal{D}(\mathcal{H}^{\otimes(i-1)}_{E})\rightarrow\mathcal{D}(\mathcal{H}^{\otimes(i-1)}_{E}\otimes\mathcal{H}_{D})$, with $\bar{E}^{i-1}=(E_{1},...,E_{i-1})$ representing all quantum states up to communication round $i$. 

    \item A decoding channel $\mathcal{P}_{B^n\to \hat{M}}$ acting on $\mathcal{H}_{B}^{\otimes n}$.
    
\end{itemize}
We denote the code by $\mathscr{C}=(\mathcal{E},\Gamma,\mathcal{P})$.

The communication scheme is illustrated in \Fig{fig:qrelay-comm-scheme}. The systems $A^n$ and $D^n$ represent the transmissions from the sender and the relay,  while $E^n$ and $B^n$ denote the received outputs at the relay and the receiver, respectively,. At the beginning of the protocol, Alice holds a ``message system" $M$, in a state $\rho_M$. Let $\ket{\Psi}\in\mathcal{D}(\mathcal{H}_{M}\otimes\mathcal{H}_{R})$ be a purification. The system $R$ represents the complementary system, to which the state in the relay communication scheme may be entangled. In the quantum communication protocol, Alice would like to  send $M$ to Bob. The encoding is done by Alice by a mapping of  $\rho_{M}\mapsto\mathcal{E}_{M\to A^{n}}(\rho_{M})$, followed by the transmission of the $A^n$ systems over $n$ channel uses.

The transmission is done iteratively, at step $i$, the relay encodes the sequence of states $(E_{i-1},\bar{E}^{i-2})$ using strictly casual encoding map $\Gamma^{(i)}_{E_{i-1}\bar{E}^{i-2}\rightarrow{\bar{E}^{i-1}D_{i}}}$, and then receives $E_{i}$ for $1\leq i\leq n$. Each step $i$ is composed of two operations, the action of $\Gamma^{(i)}$ and the action of the quantum channel $\mathcal{N}$.

The first time step $i=1$ requires an ``initial condition" for the encoding map $\Gamma^{(1)}$ as:
\begin{equation}\label{quant-rel-gamma-enc-init}
    \rho_{RA^{n}D_{1}}=\left[\left(\mathrm{id}_R\otimes \mathcal{E}_{M\rightarrow A^{n}}\right)(\Psi_{RM})\right]\otimes\Gamma_{E_{0}\rightarrow D_{1}}^{(1)}(1),
\end{equation}
where
$\Psi\equiv\ketbra{\Psi}{\Psi}$ and $E_{0}$ is a degenerate system of dimension 1. 

In what follows, we employ the notation $A_i^j \EqDef A_i A_{i+1}\ldots A_j$ for $i<j$. At $i=1$, the relay receives $E_1$ with the state, after the action of the quantum channel:
\begin{equation}\label{quant-rel-chan-init}
    \rho_{RB_{1}E_{1}A_{2}^{n}}=\left[\mathrm{id}_R\otimes \mathcal{N}_{AD\rightarrow BE}\otimes \mathrm{id}_{A_{2}^{n}}\right]\rho_{RA_{1}D_{1}A_{2}^{n}},
\end{equation}
where $A^n = A_1 A_2^n$.

The second time step $i=2$ begins with the action of $\Gamma^{(2)}$, corresponding to the relay encoding of $D_2$ as:
\begin{equation}\label{quant-rel-gamma-enc}
    \rho_{R\bar{E}_{1}D_{2}A_{2}^{n}B_{1}}=\left[\mathrm{id}_R\otimes \Gamma_{E_{1}\rightarrow \bar{E}_{1}D_{2}}^{(2)}\otimes \mathrm{id}_{A_{2}^{n}B_{1}}\right]\rho_{RB_{1}E_{1}A_{2}^{n}}.
\end{equation}

The action of the quantum channel $\mathcal{N}$ at time step $i=2$ is done by sending the state $E_2$ to the relay as:
\begin{multline}\label{quant-rel-chan}
    \rho_{RB_{2}E_{2}A_{3}^{n}B_{1}\bar{E}_{1}}=\\\left[\mathrm{id}_R\otimes \mathcal{N}_{AD\rightarrow BE}\otimes \mathrm{id}_{A_{3}^{n}B_{1}\bar{E}_{1}}\right]\rho_{RA_{2}D_{2}A_{3}^{n}B_{1}\bar{E}_{1}},
\end{multline}
and so on.

At the final time step $i=n$, the relay encodes $D_n$ as:
\begin{multline}\label{quant-rel-gamma-enc-last}
    \rho_{R\bar{E}^{n-1}D_{n}A_{n}^{n}B^{n-1}}=\\\left[\mathrm{id}_R\otimes \Gamma_{E_{n-1}\bar{E}^{n-2}\rightarrow \bar{E}^{n-1}D_{n}}^{(n)}\otimes \mathrm{id}_{A_{n}B^{n-1}}\right]\cdot\\ (\rho_{RE_{n-1}\bar{E}^{n-2}A_{n}B^{n-1}}),
\end{multline}
then the state $E_n$ is sent to the relay as the last step of the communication protocol:
\begin{multline}\label{quant-rel-chan-last}
    \rho_{RB^{n}E_{n}\bar{E}^{n-1}}=\\\left[\mathrm{id}_R\otimes \mathcal{N}_{AD\rightarrow BE}\otimes \mathrm{id}_{B^{n-1}\hat{G}_E^{n-1}}\right]\rho_{RA_{n}D_{n}B^{n-1}\hat{G}_E^{n-1}}.
\end{multline}

The output state for Bob (destination receiver) is the reduced state $\rho_{B^{n}}=\Tr_{RE_{n}\bar{E}^{n-1}}\left[\rho_{RB^{n}E_{n}\bar{E}^{n-1}}\right]$. Bob receives the channel output state $\rho_{B^{n}}$ of the output system $B^{n}$, applies the decoding channel $\mathcal{P}_{B^n\to \hat{M}}$ and recovers $\hat{M}$ as the estimation of the quantum system sent by Alice. This results in the final state $ \rho_{R\hat{M}}=\mathcal{P}_{B^n\to \hat{M}}(\rho_{RB^{n}})$. The error of the code is:
\begin{equation}\label{quant-rel-code-error}
    e^{(n)}(\Psi,\mathscr{C})=\frac{1}{2}||\Psi_{RM} - \rho_{R\hat{M}}||_{1}.
\end{equation}
A quantum transmission rate $Q>0$ is called achievable if for every 
$\varepsilon,\delta>0$ and sufficiently large $n$, there exists a $(2^{n(Q-\delta)}, n)$ code such that $e^{(n)}(\Psi,\mathscr{C})\leq \varepsilon$ for all $\ket{\Psi_{RM}}\in\mathcal{H}_M\otimes \mathcal{H}_R$. The quantum capacity $C_{\text{Q}}(\mathcal{N})$ is defined as the supremum of achievable quantum rates. 

One may focus on the task of entanglement generation. 
In this case, an entanglement-generation rate $Q_{\text{EG}}>0$ is called achievable if for every 
$\varepsilon,\delta>0$ and sufficiently large $n$, there exists a $(2^{n(Q_{\text{EG}}-\delta)}, n)$ code such that $e^{(n)}(\Phi,\mathscr{C})\leq \varepsilon$, where $\ket{\Phi_{RM}}$ is the canonical maximally entangled state on $\mathcal{H}^{\otimes 2}_M$. The entanglement-generation capacity $C_{\text{EG}}(\mathcal{N})$ is defined as the supremum of achievable entanglement-generation rates.

\section{Main Results}\label{sec:main-results}

Here we present our main results for the quantum relay channel $\mathcal{N}_{AD\rightarrow BE}$. We establish achievable rates that are based on different coding strategies with and without the entanglement assistance.

\subsection{Rate-Limited Entanglement Assistance}

Consider the three-party communication scheme in \Fig{fig:qrelay-three-party}. In the partial decode-forward strategy, the relay does not decode the entire message but only a part of it. Specifically, the quantum ``message" register $M$ consists of two subsystems, $M'$ and $M''$, with the relay decoding $M'$ alone. The corresponding transmission rates are denoted by $Q'$ and $Q''$, so that the overall rate is $Q = Q' + Q''$. 

Before stating the main results, it is important to distinguish between the two tasks for which we establish capacity bounds: quantum information transfer and entanglement generation. As outlined in \Sec{sec:introduction}, the full decode-forward strategy addresses quantum information transfer, while the partial decode-forward strategy deals with entanglement generation. The reason for this distinction is structural. In the full decode-forward case, Alice transmits a single quantum message $M$, which is fully decoded by the relay, re-encoded, and forwarded to Bob. Thus, the entire quantum information in $M$ is delivered to the destination receiver. In contrast, in the partial decode-forward case, $M$ is partitioned into two subsystems $M'$ and $M''$, and only $M'$ is decoded and re-encoded at the relay. This partial recovery and forwarding necessitates a final joint decoding at Bob, which reconstructs $M'$ and $M''$ but not the original \emph{undivided} system $M$. Consequently, the natural formulation of the partial decode-forward regime is in terms of entanglement generation rather than direct quantum information transfer. We will continue to highlight this distinction in the relevant sections and results that follow.

\subsubsection{Achievable Rates}
In the proposition below, we assume that there are pre-shared entanglement resources between Alice and Bob, and between Alice and the relay.

\begin{proposition}
\label{prop:part-dec-rates}
Consider a quantum relay channel $\mathcal{N}_{AD\rightarrow BE}$. Let $\ket{\sigma}_{A A_1 A_2 D}$ be a pure input state, and denote the corresponding output state by  $\omega_{A_1 A_2 BE}=\mathcal{N}_{AD\rightarrow BE}(\sigma_{A_1 A_2 A D})$.
Let $Q'\geq 0$ denote the rate of information that is decoded by the relay, and $Q''\geq 0$ the rate of information that is sent without the relay's help.
Assume that Alice and  Bob have access to entanglement assistance, characterized by the entanglement consumption and generation rates $L'_B,L''_B\geq 0$ and $\hat{L}'_B,\hat{L}''_B\geq 0$ in the encoding of $M',M''$, respectively. Similarly, assume that Alice and the relay have access to entanglement assistance, with consumption and generation rates $L'_E\geq 0$ and $\hat{L}'_E\geq 0$.  Then any rate point $(Q', Q'', L'_B, \hat{L}'_B, L'_E, \hat{L}'_E, L''_B, \hat{L}''_B)$ such that:
\begin{align}
\label{eq:part-dec-achievable-rates}
    &Q' + Q'' <  
        H(A_1 A_2 | D)_\sigma 
        - L'_{E} - L'_B - L''_B \nonumber\\
    &\hspace{4cm} 
        + \hat{L}'_{E} + \hat{L}'_{B} + \hat{L}''_{B}, \nonumber\\[0.6em]
    &Q' <  
        I(A_1\rangle E)_\omega 
        + L'_{E} - L'_B - \hat{L}'_{E} - \hat{L}'_{B}, \nonumber\\[0.6em]
    &Q' + Q'' <  
        I(A_1 A_2 \rangle B)_\omega 
        - L'_E + L'_B + L''_B \nonumber\\
    &\hspace{4cm} 
        - \hat{L}'_E - \hat{L}'_B - \hat{L}''_B,
\end{align}
is achievable for transmission with rate-limited entanglement assistance, where $A_1$ and $A_2$ are arbitrary auxiliary systems for the relay and direct links, respectively. 
\end{proposition}
The intepretation of the result above is given in \Sec{sec:anal-prop-1}, and
the proof for \Prop{prop:part-dec-rates} is given in \App{sec:proof-prop-1}. 

Intuitively, $A_1$ represents Alice's share of entanglement with the relay, while $A_2$ is Alice's share of entanglement with Bob. Due to entanglement monogamy, the entanglement between $A$ and $A_1$ comes at the expense of the entanglement between $A$ and $A_2$. As a result, there is a tradeoff between the entanglement that Alice generates with the relay and the entanglement that she generates with the destination decoder.

\subsection{Full Decode-Forward Bound (Unassisted)}\label{sec:full-dec-forward}

In the full decode-forward strategy, the relay fully decodes the quantum message received from Alice, re-encodes it, and forwards it to Bob. In that case, the auxiliary register $A_2$ is not needed, and the initial state is any pure state $\ket{\sigma}_{A A_1 D}$. Our result for the unassisted case is given below. Define
    \begin{equation}
    \label{eq:qdf-definition}
        \mathsf{Q}_{\text{D-F}}(\mathcal{N})\EqDef\max_{\sigma} \left[\min \left\{ I(A_1\rangle B)_\omega, I(A_1\rangle E)_\omega \right\}\right],
    \end{equation}
    s.t. $I(A_1\rangle B)_\omega\geq0$ and $I(A_1\rangle E)_\omega\geq0$, where the maximum is taken over all pure states of the form $\ket{\sigma_{A_1 A D}}$, and where $\omega_{A_1 BE}=\mathcal{N}_{AD\rightarrow BE}(\sigma_{A_1 A D})$. The subscript \mbox{`D-F'} stands for decode-forward bound. 
\begin{theorem}
\label{thm:full-dec-thm-ent-no-ass}
    The quantum capacity of the quantum relay channel $\mathcal{N}_{AD\rightarrow BE}$ without assistance satisfies:
    \begin{equation}
    C_{\text{Q}}(\mathcal{N})\geq \mathsf{Q}_{\text{D-F}}(\mathcal{N}).
\end{equation}
\end{theorem}

The proof of \Thm{thm:full-dec-thm-ent-no-ass} is given in \App{sec:proof-full-dec-no-ass}. Intuitively, the formula for in \Eq{eq:qdf-definition} exhibits a bottleneck behavior, where the worst link dictates the information throughput. 

\begin{remark}
The input state $\ket{\sigma}_{AA_1 D}$ represents the correlation that is generated between the transmitter and the relay. 
Intuitively, we think of the auxiliary system $A_1$ as Alice's entanglement resource. Due to entanglement monogamy, if the resource $A_1$ is maximally entangled with the relay's transmission, $D$, then there is no correlation with the channel input, $A$. As an extreme case, taking $\ket{\sigma}_{AA_1 D} = \ket{\psi}_A \otimes \ket{\phi_{A_1 D}}$ results in a zero rate.
Similarly, if the resource is maximally entangled with the channel input, then it has no correlation with the relay. Therefore, it is not immediately obvious how to distribute the entanglement between $A_1$, $A$, and $D$. 
\end{remark}

\begin{remark}
Typically, the direct channel from Alice to Bob is noisier than the channel from Alice to the relay, since the destination receiver is located farther away. Consider the case where Bob's channel is degraded with respect to the relay's channel. That is, there exists a degrading channel $\mathcal{D}_{E\to B}$ such that $\mathcal{N}_{AD\rightarrow B}=\mathcal{D}_{E\to B}\circ \mathcal{N}_{AD\to E}$. Based on the data processing inequality for the coherent information (see Theorem 11.9.3 in \cRef{ref:Wilde2017}), the bound reduces to 
\begin{align*}
    \mathsf{Q}_{\text{D-F}}(\mathcal{N})=
    \max_{\sigma}  I(A_1\rangle B)_\omega.
\end{align*}
\end{remark}

\begin{remark}
    In our communication setting (see \Fig{fig:qrelay-three-party}), the links are \emph{not} orthogonal; that is, $\mathcal{N}_{AD\rightarrow BE} \neq \mathcal{P}_{A\rightarrow E} \otimes \mathcal{M}_{D\rightarrow B}$. In contrast, if the channel links were orthogonal, i.e., $\mathcal{N}_{AD\rightarrow BE} = \mathcal{P}_{A\rightarrow E} \otimes \mathcal{M}_{D\rightarrow B}$,  \Thm{thm:full-dec-thm-ent-no-ass} yields:
    \begin{equation*}
        \mathsf{Q}_{\text{D-F}}(\mathcal{N}) \geq \min \left\{ I_{c}(\mathcal{M}), I_{c}(\mathcal{P}) \right\},
    \end{equation*}
    in the unassisted case.
    The orthogonal links model was previously considered by Ghalaii and Pirandola\cc{ref:Ghalaii2020, ref:Pirandola2016}, Gyongyosi and Imre\cc{ref:Gyongyosi2012, ref:Gyongyosi2014}, and Shi et al.\cc{ref:Shi2012}. Intuitively, the generalized communication model allows for entanglement between the inputs of Alice and the relay, allowing for a higher achievable rate.
\end{remark}

\subsection{Partial Decode-Forward Bound (Unassisted)}\label{sec:part-dec-forward}

The main result of our work is given below. Define
    \begin{multline}
    \label{eq:qpdf-definition}
        \mathsf{Q}_{\text{PD-F}}(\mathcal{N})\EqDef \max_{\sigma} \bigl[ \min \bigl\{ I(A_1 A_2\rangle B)_\omega \,,\\\; I(A_1\rangle E)_\omega  + I(A_2\rangle B  A_1)_\omega \bigr\}\bigr]
    \end{multline}
      where the maximum is taken over all pure states of the form $\ket{\sigma_{A_1 A_2 A D}}$ s.t. $I(A_1\rangle B)_\omega\geq0$, $I(A_1\rangle E)_\omega\geq0$ and $I(A_2\rangle B  A_1)_\omega\geq0$, and where $\omega_{A_1 A_2 BE}=\mathcal{N}_{AD\rightarrow BE}(\sigma_{A_1 A_2 A D})$. The subscript \mbox{`PD-F'} stands for partial decode-forward bound. 
\begin{theorem}
\label{thm:part-dec-thm-ent-no-ass}
    The entanglement-generation capacity of the quantum relay channel $\mathcal{N}_{AD\rightarrow BE}$ without assistance satisfies:
    \begin{equation}
    C_{\text{EG}}(\mathcal{N})\geq \mathsf{Q}_{\text{PD-F}}(\mathcal{N}).
\end{equation}
\end{theorem}

The proof of \Thm{thm:part-dec-thm-ent-no-ass} is given in \App{sec:proof-part-dec-no-ass}.

\begin{remark}
\Thm{thm:part-dec-thm-ent-no-ass} can be interpreted as follows. 
As in Proposition~\ref{prop:part-dec-rates}, the ``message" register is split into two components, $M'$ and $M''$,
at quantum rates $Q'$ and $Q''$, with a total entanglement-generation rate of $Q = Q' + Q''$, as defined in \Prop{prop:part-dec-rates}. The relay only decodes and encodes the first component, whereas the second component is decoded by Bob without the relay's help.
The rate $Q'$ is chosen as
\begin{equation*}
    Q' =\min \left\{ I(A_1\rangle B)_\omega, I(A_1\rangle E)_\omega \right\},
\end{equation*}
in a similar manner as in the full decode-forward bound (see
\Eq{eq:qdf-definition}), 
and the rate $Q''$ as
\begin{equation*}
    Q'' = I(A_2 \rangle B A_1)_\omega.
\end{equation*}
The expression $I(A_1 A_2 \rangle B)_\omega$ in \Eq{eq:qpdf-definition} can be recovered from the entropy chain rule. 
\end{remark}

\begin{remark}
Intuitively, the information rate at the relay $Q'$ is determined by the bottleneck, i.e., the weaker link on the relay path. If the Alice-to-relay link is noisier than the relay-to-Bob link, then $I(A_1 \rangle E)_\omega$ limits $Q'$. Whereas, if the relay-to-Bob channel is noisier, the limiting term becomes $I(A_1 \rangle B)_\omega$. 
%
The formula for $Q''$ does not involve the relay output $E$, since the relay is not required to decode this part of the information.  
While $A_1$ cannot be copied due to the no-cloning theorem, it \emph{effectively} serves as side information at the receiver when decoding $A_2$.
Indeed, 
using entropy identities, we can write
\begin{equation*}
    I(A_2 \rangle B A_1)_\omega = I(A_2 \rangle B \mid A_1)_\omega,
\end{equation*}
which can be interpreted as Bob decoding the message component $M''$ ``given" the message system $M'$ sent via the relay.

\end{remark}

\subsection{Entanglement-Assisted Capacity}

In the next theorem, we consider communication of quantum information via the quantum relay channel under the full decode-forward strategy, assuming that Alice and Bob are provided with pre-shared and unlimited entanglement assistance. In this strategy, the relay fully decodes the quantum message, re-encodes it, and forwards it, hence, the auxiliary register $A_2$ is not required, and the initial state can be taken as any pure state $\ket{\sigma}_{A A_1 D}$ (see also \Sec{sec:full-dec-forward}). We denote the entanglement-assisted quantum capacity by $C_{\text{Q-EA}}(\mathcal{N})$, where the maximization is as before, and the subscript `Q-EA' indicates the transmission of quantum information using entanglement assistance.  This can be viewed as teleportation via a noisy relay channel \cite[Sec. 1.3.7]{nielsen2010quantum}. 

Let
   \begin{equation*}
        \mathsf{Q}_{\text{EA,D-F}}(\mathcal{N})\EqDef\max_{\sigma} \left[\frac{1}{2}I(A_1 ; B)_\omega - \frac{1}{2}I(A_1 ; D)_\sigma\right].
    \end{equation*}

\begin{theorem}
\label{thm:full-dec-thm-ent-with-ass} 
    The entanglement-assisted quantum capacity of the quantum relay channel $\mathcal{N}_{AD\rightarrow BE}$ satisfies:
    \begin{equation}
    C_{\text{Q-EA}}(\mathcal{N})\geq \mathsf{Q}_{\text{EA,D-F}}(\mathcal{N}).
\end{equation}
\end{theorem}

The proof of \Thm{thm:full-dec-thm-ent-with-ass} is given in \App{sec:proof-part-dec-with-ass}. In the analysis, we take  $L'_B > \hat{L}'_B > 0$ and $L''_B > \hat{L}''_B > 0$ with $L'_E = \hat{L}'_E \to 0$ in \Prop{prop:part-dec-rates}.

\begin{remark}
The theorem above has an immediate consequence on the transmission of classical information with entanglement assistance. Specifically, using the superdense coding protocol \cite[Sec. 2.3]{nielsen2010quantum}, classical information can be sent at a rate of $R\geq$ $ 2\cdot \mathsf{Q}_{\text{EA,PD-F}}(\mathcal{N})$ (classical information bits per transmission).
\end{remark}

\begin{remark}
    The results bear resemblance to the entanglement-assisted quantum capacity of a quantum channel with side information at the transmitter, as studied in \cRef{ref:Dupuis2009}. In the side information model, the channel $\tilde{\mathcal{N}}_{DA\to B}$ is governed by an external environment $D$, which is entangled with the transmitter. The connection becomes apparent if we interpret the transmission to the relay as entangling the transmitter with the channel's environment (see \Fig{fig:qrelay-three-party}). Note, however, that  the environment state in the side information model\cc{ref:Dupuis2009} is fixed, i.e., $\sigma_D$ is dictated by the model. In contrast, our formula involves optimization as we are free to choose the entanglement constellation between the transmitter and the relay. 
\end{remark}

In the upcoming sections, we provide interpretation and a concrete example to further illustrate the results for the achievable rates in \Thm{thm:full-dec-thm-ent-no-ass}, \Thm{thm:part-dec-thm-ent-no-ass} and \Thm{thm:full-dec-thm-ent-with-ass}.

\section{Interpretation}
\label{sec:anal-prop-1}

In what follows we explain the intuition behind each of the three conditions given by \Eq{eq:part-dec-achievable-rates} in \Prop{prop:part-dec-rates}.

\paragraph{Full Decode-Forward Interpretation}

First, suppose that the relay decodes the entire message. In this case, we set $Q'' \to 0$ and $L''_B = \hat{L}''_B \to 0$. We also set $A_2$ to be degenerate (say, a register of dimension $1$). Next, we define the net entanglement consumption of EPR pairs as $\Delta L'_B = L'_B - \hat{L}'_{B}$ and $\Delta L'_E = L'_E - \hat{L}'_{E}$,  shared between Alice and Bob and between Alice and the relay, respectively. Under these assumptions, \Eq{eq:part-dec-achievable-rates} yields
\begin{equation}
\label{eq:full-dec-rates-intuition}
\begin{split}
    &Q' + \Delta L'_B + \Delta L'_E \leq H(A_1 |D)_\sigma\\
    &Q' - \Delta L'_E \leq I(A_1 \rangle E)_\omega - (L'_B + \hat{L}'_{B})\\
    &Q' - \Delta L'_B \leq I(A_1 \rangle B)_\omega - (L'_E + \hat{L}'_{E}).
\end{split}
\end{equation}
Then, the first condition in \Eq{eq:full-dec-rates-intuition} states that the quantum information transmitted by Alice, together with the entanglement consumed from her share of the EPR pairs, must fit within the channel input available to Alice, conditioned on the relay's input. 

In a Block Markov coding scheme,  Alice transmits a sequence of blocks $A^n(1),\ldots,A^n(T)$, each of size $n$, in order to distribute a ``quantum message" sequence $M(2),\ldots,M(T)$, each of dimension $2^{nQ}$. The transmission in each block not only carries new information but also forwards part of the information from the previous block. 


The second condition in \Eq{eq:full-dec-rates-intuition} bounds the net information rate $Q'-\Delta L_E'$, having accounted for the entanglement resources with the relay. The subtraction of  $(L'_B+\hat{L}'_{B})$ reflects the bottleneck introduced by the relay. This arises due to the nature of our Block Markov coding scheme, where the entanglement assistance available to Bob in block $j$ is effectively generated by the relay in block $j-1$.  


The third condition shown in \Eq{eq:full-dec-rates-intuition} has a similar structure. Here, we subtract $(L'_E+\hat{L}'_{E})$, again reflecting the relay bottleneck. In addition, $\Delta L'_B$ has a negative sign (opposite to the first condition), corresponding to the net entanglement consumption between Alice and Bob.

\paragraph{Partial Decode-Forward Interpretation}
The above discussion generalizes to the case where part of the information is sent without the relay's help, i.e., $Q'' > 0$. Recall that our coding scheme includes two message registers $M'$ and $M''$, of rates $Q'$ and $Q''$, respectively, as presented in \Eq{eq:part-dec-achievable-rates}. The relay decodes $M'$ alone, and does not decode $M''$. In this case, we define $\Delta L''_B = L''_B - \hat{L}''_{B} $ as the net entanglement consumption in the encoding of $M''$. 

Based on the intuition given above for the \Eq{eq:full-dec-rates-intuition}, the \Eq{eq:full-dec-rates-intuition} changes to
\begin{equation}
\label{eq:part-dec-rates-intuition}
\begin{split}
    &Q' + \Delta L'_B + \Delta L'_E + \Delta L''_B \leq H(A_1 |D)_\sigma\\
    &Q' - \Delta L'_E \leq I(A_1 \rangle E)_\omega - (L'_B + \hat{L}'_{B})\\
    &Q' - \Delta L'_B - \Delta L''_B \leq I(A_1 \rangle B)_\omega - (L'_E + \hat{L}'_{E}).
\end{split}
\end{equation}
where only the first and third conditions now include the additional term $\Delta L''_B$, accounting for the entanglement consumed for the encoding of $M''$, and the net entanglement consumption in the decoding of $M''$, respectively.

\section{Example: Quantum erasure channel}\label{sec:quant-relay-examples}

\begin{figure}[t]
  \centering{\includegraphics[width=\linewidth]{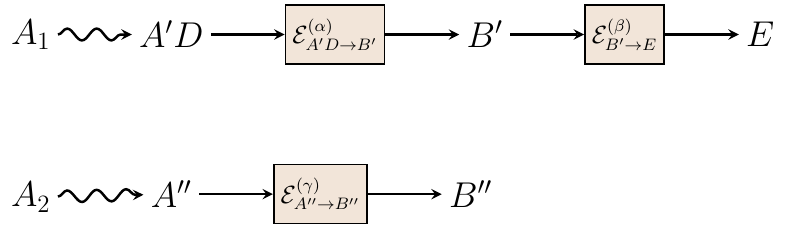}}
   \caption{A quantum relay channel with orthogonal components considered in our example. Alice's (sender's) input $A$ splits into $A'$ and $A''$, entangled with auxiliaries $A_1$ and $A_2$. $A'$ is sent through a relay-assisted link, while $A''$ is sent directly to Bob (receiver). The receiver obtains the outputs $B'$ and $B''$, respectively.}
    \label{fig:qrelay-example-erasure}
\end{figure}

To illustrate our results, we consider the quantum erasure channel, a widely used model in quantum communication\cc{ref:Grassl1997, ref:Bennett1997}. An erasure channel with erasure probability $p$ either transmits the input quantum state perfectly with probability $1-p$, or replaces it with an ``erasure'' state $\ket{e}$ with probability $p$. Formally, for an input state $\rho$ on a qudit space $\mathbb{C}^d$ of dimension $d$, the channel acts as
\[
\mathcal{E}^{(p)}(\rho) = (1-p)\,\rho + p\,\ket{e}\!\bra{e},
\]
where $\ket{e}$ is orthogonal to the qudit space $\mathbb{C}^d$. Hence, the output dimension is $d+1$.

We consider a quantum relay channel that consists of orthogonal components, as shown in \Fig{fig:qrelay-example-erasure}. Specifically, Alice's input $A$ consists of two components, $A'$ and $A''$, which are sent through separate links. The first link is repeater aided and goes through  the relay, whereas the latter is a direct repeaterless link to Bob. The system at the destination receiver, Bob, consists of the respective outputs, $B'$ and $B''$. We assume that $A',A'', D, B',B'', E$ are all single qubit registers. The overall quantum channel is
\begin{equation}
    \mathcal{N}_{A'A''D\to B'B''E} = (\mathcal{E}^{(\beta)}_{B'\to E}\circ\mathcal{E}^{(\alpha)}_{A'D\to B'})\otimes\mathcal{E}^{(\gamma)}_{A''\to B''}.
\end{equation}
Naturally, we choose the ancilla $A_1$ to be maximally entangled with the relay-link input, $A'$, and the ancilla $A_2$ to be maximally entangled with the direct-link input, $A''$. The initial state is thus taken as $\ket\sigma=\ket{\Phi_{A_1 A' D}}\otimes\ket{\Phi_{A_2 A''}}$, where $A_1$ is a two-qubit register and $A_2$ is a single-qubit register.

Now we proceed to calculate the coherent information quantities under the partial decode-forward bound given in \Thm{thm:part-dec-thm-ent-no-ass}. For the coherent information to Bob, we have by the data processing inequality for the coherent information (see Theorem 11.9.3 in \cRef{ref:Wilde2017}): 
\begin{equation}
    I(A_1\rangle B' B'')_\omega\geq I(A_1\rangle B') = (1-2\alpha)\log_{2}4,
\end{equation}
and for the coherent information to the relay,  
\begin{equation}
   I(A_1\rangle E)_\omega = (1-2q)\log_{2}4,
\end{equation}
with $q=\alpha+(1-\alpha)\beta$. Finally, the ``conditional term" is bounded as 
\begin{equation}
   I(A_2\rangle B' B'' A_1) \geq I(A_2\rangle B'') = 1-2\gamma.
\end{equation}
Considering  the partial decode-forward  rate  in \Eq{eq:qpdf-definition}, we obtain:
\begin{equation}
   \mathsf{Q}_{\text{PD-F}}(\mathcal{N}) \geq \min \bigl\{ (1-2\alpha)\log_{2}4 \,,\\\; (1-2q)\log_{2}4 \bigr\} + (1-2\gamma).
\end{equation}
We note that $q\geq \alpha$, hence the first term reduces to $2(1-2q)$.

Based on Theorem~\ref{thm:part-dec-thm-ent-no-ass}, the entanglement-generation capacity of the quantum relay channel above thus satisfies
\begin{equation}
   C_{\text{EG}}(\mathcal{N}) \geq 2(1-2q) + (1-2\gamma),
\end{equation}
and we can see that we benefit from dividing the transmission between the repeater-aided and the repeaterless links. 

\section{Summary and Discussion}\label{sec:summary-conclusion}

We study quantum information transmission over a three-terminal quantum relay channel, focusing on decode-forward coding strategies, in both entanglement-assisted and unassisted settings, with either full or partial decoding at the relay. Our model captures key tradeoffs between repeater-aided and repeaterless communications, and highlights the role of entanglement between the transmitter and the relay. Furthermore, our framework generalizes previous results, which focus on either repeaterless transmission or repeater-aided communication separately\cc{ref:Pirandola2019, ref:Winnel2021, ref:Buml2015, ref:Dupuis2009, ref:Pereg2024-Relay}. These insights serve as a stepping stone towards the design of quantum networks (in the long-term, a quantum Internet), and provide a foundation for practical quantum repeaters and long-distance QKD.

One avenue for future research is to develop alternative coding strategies. In particular, it would be interesting to explore schemes similar to those considered in \cRef{ref:Pereg2024-Relay}, such as measure-forward, assist-forward, or compress-and-forward \cc{ref:Wu2013}, extending beyond the partial and full decode-forward strategies. Investigating these approaches for quantum channels with quantum messages could reveal new trade-offs between relay-assisted and direct transmission, potentially improving achievable rates or reducing entanglement consumption. Another promising direction is to apply our results to bosonic channels \cc{ref:Ghalaii2022, ref:Giovannetti2004}, which model optical communication systems. Extending partial and full decode-forward strategies to bosonic modes could provide insight into achievable rates in realistic continuous-variable quantum networks, including fiber-optic and free-space links.



Another interesting direction is to explore the relationship between quantum relaying and superactivation. 
 Both phenomena demonstrate that the communication capabilities of quantum networks can exceed what might be inferred from the constituent channels in isolation.
Establishing a concrete relationship between these two mechanisms may lead to new insights into the structure of quantum network capacities.

On a more practical side, when considering how quantum relay channels can be used in the context of distributed relays that facilitate high-rate quantum error correction, it is interesting to analyze our results for quantum channels widely used to model noise in quantum hardware\cc{ref:Ilin2024, ref:Ilin2025}. For example, even local Markovian noise models, such as non-uniform depolarization, generalized amplitude damping, and coherent two-qubit error models often arising from controlled-X gates used in variational quantum algorithms, may provide interesting insights\cc{ref:Ilin2024, ref:Ilin2025, ref:Fontana2021}. Analyzing these channels would bridge our primarily quantum communication-focused results with the practical realm of quantum computation.

\section*{Acknowledgments}
The authors thank Gerhard Kramer (TUM)
for useful discussions.
Yilin and Pereg were supported by  Israel Science Foundation (ISF), 
 Grants n. 939/23 and 2691/23,
Deutsche Forschungsgemeinschaft (DFG) through the
 German-Israeli Project Cooperation (DIP) 
 n. 2032991, Ollendorff-Minerva Center of the Technion 
 n. 86160946,   
 Israel Innovation Authority Program for  Post-Quantum Communications n. 2035868,
  the Planning and Budgeting Committee of the Council for Higher Education of Israel
(VATAT) for the QERNEL Quantum  Computing Research Hub n.
2072651, and the  Helen Diller Quantum Center at the Technion.
Pereg was also supported by Chaya Chair n. 8776026.

\begin{appendices} %
{

\section{Analytic Tools: The FQSW Theorem}\label{sec:fqsw-thrm}

Here we give a quick overview of the fully quantum Slepian-Wolf (FQSW) protocol \cc{ref:Dupuis2009, ref:Abeyesinghe2009}. Suppose we have a pure state $\ket{\psi}_{ABC}$ shared by Alice, Bob and some reference system Charlie. Alice wishes to sent her state to Bob with high fideluty and using as few qubits as possible. The FQSW protocol states that Alice first applies some unitary $U_{A\rightarrow A_1 A_2}$, for example drawn from the Haar measure, and then sending $A_2$ to Bob.

In the fully quantum FQSW protocol, the key requirement is that after Alice applies a random unitary $U_{A \rightarrow A_1 A_2}$, the subsystems $A_1$ and $C$ are left in a product state. Since Bob holds the purifying system of $A_1 C$, there must exist a local unitary $U_B$ on his side that transforms his purification into two separate purifying systems: one for $A_2$ and one for $C$. The purification of $C$ is the composite system $A_2 B$, so that the system $A_2$ sent by Alice, together with Bob's register $B$, recovers the original correlations. Meanwhile, $A_1$ is paired with its purification to form an EPR state shared between Alice and Bob. Therefore, in addition to transmitting $A_2$, the by-product of the FQSW protocol is an entanglement resource shared by Alice and Bob produced at the end of the protocol.


As shown in \cRef{ref:Dupuis2009}, it is possible to calculate how close is a sub-system on $A_1C$ is in a product state. Averaging over all choices of $U_{A\rightarrow A_1 A_2}$ yields:
\begin{equation}
\label{eq:quant-rel-FQSW-thrm}
    \int_{\mathcal{U}(A)}\left\|\rho_{A_1 C}-\frac{\Id_{A_1}}{|A_1|}\otimes\psi_C\right\|^2_1 dU \leq \frac{|A||C|}{|A_2|^2}\Tr\left[\left(\psi_{AC}\right)^2\right].
\end{equation}
Performing this protocol over $n$ transmissions from Alice to Bob for a very large $n$ leads to $\Tr\left[\left(\psi_{AC}\right)^2\right]$ converging to $2^{-nH(AC)}$, where $H(AC)$ is the von Neumann entropy of the reduced state $\psi_{AC}$ of the initial pure state $\ket{\psi}_{ABC}$. The latter is a consequence of the quantum asymptotic equipartition property, that for large $n$, the state restricted to the typical subspace is approximately maximally mixed, so its purity is determined by the von Neumann entropy\cc{ref:Tomamichel2009-QAEP}. Note that \Eq{eq:quant-rel-FQSW-thrm} holds on average, therefore, there must exist at least one unitary that satisfies \Eq{eq:quant-rel-FQSW-thrm}.

Based on the discussion above, we write
\begin{equation}
    \rho_{A_1 C}\fqswEpsilon{\varepsilon}\frac{\Id_{A_1}}{|A_1|}\otimes\psi_C
\end{equation}
to indicate that the two states are close in trace distance, i.e.,
\begin{equation}
    \frac{1}{2}\left\|\rho_{A_1 C}-\frac{\Id_{A_1}}{|A_1|}\otimes\psi_C\right\|_1\leq\varepsilon,
\end{equation}
for some random unitary $U_{A\rightarrow A_1 A_2}$, 
with the proximity parameter $\varepsilon = \frac{|A||C|}{|A_2|^2}2^{-nH(AC)}\ll 1$.}

{
\section{Proof of Proposition~\ref{prop:part-dec-rates}}\label{sec:proof-prop-1}

First, we introduce the notation for the entanglement resources consumed and generated during the protocol. For communication between the transmitter and the destination receiver via the relay, the quantum systems $G'_E$ and $G'_B$ represent the pre-shared entanglement resources available at the relay and the destination receiver (Bob), respectively, with associated consumption rates defined as $L'_{E} \EqDef \frac{1}{n} \log |G'_E|$ and $L'_{B} \EqDef \frac{1}{n} \log |G'_B|$. For direct communication between the transmitter and the destination receiver, the quantum system $G''_B$ represents the pre-shared entanglement resource available at the destination receiver (Bob) with associated consumption rate $L''_{B} \EqDef \frac{1}{n} \log |G''_B|$.

In accordance with \App{sec:fqsw-thrm}, for communication between the transmitter and the destination receiver via the relay, the quantum systems $\hat{G}'_{1,A}$ and $\hat{G}'_{2,A}$ represent the by-product entanglement generated between the transmitter (Alice) and the relay, and between the transmitter and the destination receiver, respectively. These systems reside at the transmitter, with corresponding generation rates defined as $\hat{L}'_{E} \EqDef \frac{1}{n} \log |\hat{G}'_{1,A}|$ and $\hat{L}'_B \EqDef \frac{1}{n} \log |\hat{G}'_{2,A}|$. For direct communication between the transmitter and the destination receiver, the quantum system $\hat{G}''_{A}$ represents the by-product entanglement generated between the transmitter and the destination receiver, with an associated generation rate of $\hat{L}''_{B} \EqDef \frac{1}{n} \log |\hat{G}''_{A}|$.

\subsection{Decoupling Properties}
We begin with decoupling properties. Let $\ket{\sigma_{A_1 A_2 A D}}$. We use rate splitting. That is, our quantum ``message system" $M$ consists of two components, $M'$ and $M''$, where $M'$ is the message  decoded and then sent  by the relay, and $M''$  is the message component sent directly to the destination receiver (Bob). The overall quantum rate is $Q=Q'+Q''$, as defined in \Prop{prop:part-dec-rates}, where the rates $Q'$ and $Q''$ are associated with the dimensions of $M'$ and $M''$, respectively.
The relay part, $M'$, is encoded into $A_1^n$, which in turn is  split into 
$\{R', G'_E, G'_B, \hat{G}_{1, A}, \hat{G}_{2, A}\}$. Similarly, the direct-transmission component, $M''$, is encoded into $A_2^n$, and then split into $\{R'', G''_B, \hat{G}''_{A}\}$.

\subsubsection{Decoupling at the encoder}\label{sec:init-state-split}

Consider the product state, $\ket{\sigma_{A_1 A_2 A D}}^{\otimes n}$, where $A_1$ and $A_2$ are the encoder ancillas, \mbox{$A$ is} the channel input, and $D$ is the relay output. Projecting onto the typical subspaces of $A_1$ and $A_2$ gives:
\begin{equation}
    \ket{\psi_{A_{1,\text{typ}}^n A_{2,\text{typ}}^n A^n D^n}} = (\Pi_{A_1^n A_2^n \to A_{1,\text{typ}}^n A_{2,\text{typ}}^n} \otimes \Id) \ket{\sigma_{A_1 A_2 A D}}^{\otimes n}.
\end{equation}
As in \App{sec:fqsw-thrm}, we apply  random unitaries $U_{A_{1,\text{typ}}^n \to R' G'_E G'_B \hat{G}'_{1, A} \hat{G}'_{2, A}}$ and $U_{A_{2,\text{typ}}^n \to R'' G''_B \hat{G}''_{A}}$, where $R'$ is the reference system with $|R'|=2^{nQ'}$ for the relay link and $R''$ is the reference system with $|R''|=2^{nQ''}$ for the direct link. The resulting state is
\begin{multline}
\label{eq:theta-state-init-state}
    \ket{\theta_{R' G'_E G'_B \hat{G}'_{1, A} \hat{G}'_{2, A} R'' G''_B \hat{G}''_{A} A^n D^n}} =\\ \left(U_{A_{1,\text{typ}}^n} \otimes U_{A_{2,\text{typ}}^n} \otimes \Id\right) \ket{\psi_{A_{1,\text{typ}}^n A_{2,\text{typ}}^n A^n D^n}}.
\end{multline}
Considering the reduced state $\theta_{R' G'_E G'_B R'' G''_B D^n}$, we can ``decouple'' $D^n$ from $R' G'_E G'_B R'' G''_B$, i.e.,
\begin{equation}
    \theta_{R' G'_E G'_B R'' G''_B D^n} \fqswEpsilon{\varepsilon_1} \frac{\Id_{R' G'_E G'_B R'' G''_B }}{|R' G'_E G'_B R'' G''_B|} \otimes \sigma_{D^n},
\end{equation}
which, by the FQSW Theorem (see \App{sec:fqsw-thrm}), holds for
\begin{align}
    &\varepsilon_1\equiv \varepsilon_1(n) = \frac{|A_{1,\text{typ}}^n| \cdot |A_{2,\text{typ}}^n| \cdot |D^n|}{|\hat{G}'_{1, A}|^2 |\hat{G}'_{2, A}|^2 |\hat{G}''_{A}|^2} \cdot 2^{-n H(A_1 A_2 D)_\sigma}
    \nonumber\\
    &\leq 
    2^{-n(H(A_1 A_2 | D)_\sigma - L'_{E} - L'_B - L''_B + \hat{L}'_{E} + \hat{L}'_{B} + \hat{L}''_{B} - Q' - Q'' -\delta)},
\end{align}
and sufficiently large $n$. The last inequality follows from standard typical projector properties (see Property 15.1.2 in \cRef{ref:Wilde2017}).
Observe that $\varepsilon_1(n)$ tends to zero as $n\to\infty$, provided that 
\begin{equation}
\label{eq:init-state-split-Q-ass}
    Q' + Q'' < H(A_1 A_2 | D)_\sigma - L'_{E} - L'_B - L''_B + \hat{L}'_{E} + \hat{L}'_{B} + \hat{L}''_{B} - \delta.
\end{equation}

\subsubsection{Decoupling at the relay}\label{sec:rel-state-split}

Consider a Stinespring representation of the quantum relay channel $\mathcal{N}_{AD\to BE}$, with an isometry $V^{\mathcal{N}}_{AD\to B E J_O}$. Applying this isometry produces the following output state:
\begin{multline}
\label{eq:decoup-chan-output}
    \ket{\phi_{A_{1,\text{typ}}^n A_{2,\text{typ}}^n B^n E^n J_O^n}} =\\ (\Id\otimes (V^{\mathcal{N}}_{AD\to B E J_O})^{\otimes n})\ket{\theta_{A_{1,\text{typ}}^n A_{2,\text{typ}}^n A^n D^n}},
\end{multline}
where $E^n$ is received at the relay, $B^n$ at  the destination receiver, and $J_O^n$ represents the environment of the joint output, $B^n E^n$. To proceed, recall that $A_{1,\text{typ}}^n \to R' G'_E G'_B \hat{G}'_{1, A} \hat{G}'_{2, A}$. We denote the composite environment of the relay by $J_E \EqDef B J_O$, and consider the reduced state $\phi_{R' \hat{G}'_{1, A} \hat{G}'_{2, A} G'_B J_E^n}$. Decoupling the relay's environment:
\begin{equation}
    \phi_{R' \hat{G}'_{1, A} \hat{G}'_{2, A} G'_B J_E^n} \fqswEpsilon{\varepsilon_2} \frac{\Id_{R' \hat{G}'_{1, A} \hat{G}'_{2, A} G'_B}}{|R' \hat{G}'_{1, A} \hat{G}'_{2, A} G'_B|} \otimes (\mathcal{N}_{A\to J_E}(\sigma_A))^{\otimes n},
\end{equation}
where
\begin{align}
    \varepsilon_2(n) &= \frac{|A_{1,typ}|\cdot|J_E^n|}{|G'_E|^2} \cdot 2^{-nH(A_1 J_E)_\omega}\nonumber\\
    &\leq 
    2^{-n(I(A_1 \rangle E)_\omega + L'_{E} - L'_B - \hat{L}'_{E} - \hat{L}'_{B} - Q' - \delta)},
\end{align}
for sufficiently large $n$, where the inequality follows from the same arguments as used in \App{sec:init-state-split}. The last expression tends to zero as $n\to\infty$, if
\begin{equation}
\label{eq:relay-state-split-Q-ass}
    Q' < I(A_1\rangle E)_\omega + L'_{E} - L'_B - \hat{L}'_{E} - \hat{L}'_{B} - \delta,
\end{equation}
with respect to $\omega_{A_1 A_2 BE}=\mathcal{N}_{AD\rightarrow BE}(\sigma_{A_1 A_2 A D})$.

\subsubsection{Decoupling at the destination receiver}\label{sec:receive-state-split}

We now consider the overall transmission from Alice to Bob, treating the relay’s output as part of Bob’s environment. The output state is as in \Eq{eq:decoup-chan-output}. Recall that $A_{1,\text{typ}}^n \to R' G'_E G'_B \hat{G}'_{1, A} \hat{G}'_{2, A}$ and $A_{2,\text{typ}}^n \to R'' G''_B \hat{G}''_{A}$. Denote Bob’s environment by $J_B \EqDef E J_O$ and consider the reduced state $\phi_{R' \hat{G}'_{1, A} \hat{G}'_{2, A} G'_E R'' \hat{G}''_A J_B^n}$. Similarly as before, decoupling  Bob's environment yields
\begin{multline}
    \phi_{R' \hat{G}'_{1, A} \hat{G}'_{2, A} G'_E R'' \hat{G}''_A J_B^n} \fqswEpsilon{\varepsilon_3}\\ \frac{\Id_{R' \hat{G}'_{1, A} \hat{G}'_{2, A} G'_E R'' \hat{G}''_A}}{|R' \hat{G}'_{1, A} \hat{G}'_{2, A} G'_E R'' \hat{G}''_A|}\otimes (\mathcal{N}_{A\to J_B}(\sigma_{A}))^{\otimes n},
\end{multline}
where
\begin{align}
\varepsilon_3(n) &= \frac{|A_{1,\mathrm{typ}}| \cdot |A_{2,\mathrm{typ}}|\cdot|J_B^n|}{|G'_B|^2|G''_B|^2}\cdot 2^{-nH(A_1 A_2 J_B)_\omega}\nonumber\\
    &\leq 
    2^{-n(I(A_1 A_2 \rangle B)_\omega - L'_E + L'_B + L''_B - \hat{L}'_E - \hat{L}'_B - \hat{L}''_B - Q' - Q'' - \delta)},
\end{align}
for sufficiently large $n$. This tends to zero if
\begin{equation}
\label{eq:receive-state-relay-split-Q-ass}
    Q' + Q'' < I(A_1 A_2 \rangle B)_\omega - L'_E + L'_B + L''_B - \hat{L}'_E - \hat{L}'_B - \hat{L}''_B - \delta.
\end{equation}

\subsection{Encoding and decoding schemes}
\label{sec:encoding-decoding-isometries}

In this section, we describe the encoding and decoding procedures for the sender, relay, and destination receiver. We use a Block Markov scheme: Alice transmits a sequence of $T$ blocks $A^n(1),\ldots,A^n(T)$ each block is of size $n$, in order to distribute $M(2),\ldots,M(T)$, each of dimension $2^{nQ}$. The effective transmission rate is thus $\frac{T-1}{T}Q$, which tends to $Q$ in the limit of $T\to\infty$.

Throughout, we use the following convention for block indexing to clearly distinguish between the current and past instances of a quantum system during the $T$-block encoding and decoding processes. A quantum system labeled as $S$ refers to the system at block index $j$, i.e., $S \EqDef S(j)$. A quantum system labeled as $\ubar{S}$ refers to the system at the previous block, i.e., $\ubar{S} \EqDef S(j-1)$.

As outlined in \Sec{sec:main-results}, our results focus on two strategies: the full decode-forward coding strategy for the transmission of quantum information, and the partial decode-forward coding strategy for entanglement generation. The encoding and decoding schemes described below apply to both cases, with one key distinction. 

For the full decode-forward strategy, at the beginning of the protocol Alice holds a message system $M$, with an arbitrary purification $\ket{\Psi}_{RM}$ 
(see also \Sec{sec:coding-scheme}). In contrast, in the partial decode-forward coding scheme, Alice prepares a product of maximally entangled states $\ket{\Phi}_{R'M'} \otimes \ket{\Phi}_{R''M''}$, where $\ket{\Phi}$ denotes the canonical maximally entangled state, and where the joint reference system $R'R''$ represent Alice's entanglement resource, while the  message system $M'M''$ represents the entanglement resource that is sent to Bob. 

Accordingly, in what follows we present the encoding and decoding schemes for the partial decode-forward case, using the notation $\Phi = \ketbra{\Phi}{\Phi}$ for purifications of Alice’s message systems (see also \Sec{sec:coding-scheme}). The adaptation to the full decode-forward case is obtained by replacing $\Phi \to \Psi$, and by accounting for the fact that the relay now decodes and re-encodes the entire message (see \Sec{sec:full-dec-forward} and \Sec{sec:part-dec-forward} for further discussion).

\subsubsection{Transmitter, encoder and channel isometries}

In order to obtain the transmitter and encoder isometries, 
we make use of Uhlmann’s theorem as stated in \Sec{sec:preliminaries-notation}.

For each block $j$, there exists a family of partial isometries $W_{M' M'' G'_A G''_A \to A^n \hat{G}'_{1,A} \hat{G}'_{2,A} \hat{G}''_A}$ such that, according to the \App{sec:init-state-split}, the following approximation holds:
\begin{multline}
    \ket{\theta_{R' G'_E G'_B \hat{G}'_{1, A} \hat{G}'_{2, A} R'' G''_B \hat{G}''_{A} A^n D^n}} \fqswEpsilon{4\sqrt{\varepsilon_1}}\\
    \left(W_{M' M'' G'_A G''_A \to A^n \hat{G}'_{1,A} \hat{G}'_{2,A} \hat{G}''_A} \otimes \Id \right)\cdot\\
    \bigl( \Phi_{R'M'} \otimes\Phi_{R''M''} \otimes\Phi_{G'_{1,A} G'_E} \otimes\\ \Phi_{G'_{2,A} G'_B} \otimes \Phi_{G''_A G''_B} \otimes \sigma_{D^n} \bigr),
\end{multline}
where $\Phi_{R'M'}$ and $\Phi_{R''M''}$ denote the entangled "message states" for the relay and direct link, with reference systems $R', R''$ and message registers $M', M''$, respectively. 
The product of maximally entangled states $\Phi_{G'_{1,A} G'_E} \otimes \Phi_{G'_{2,A} G'_B} \otimes \Phi_{G''_A G''_B}$ represents the entanglement assistance shared between Alice and the relay, and between Alice and Bob, for both links. The output systems $A^n \hat{G}'_{1,A} \hat{G}'_{2,A} \hat{G}''_A$ jointly purify the reduced state $\theta_{R' R'' G'_E G'_B G''_B D^n}$.

The channel output is then
\begin{multline}
    \ket{\phi_{A_{1,\text{typ}}^n A_{2,\text{typ}}^n B^n E^n J_O^n}} =\\ (\Id\otimes (V^{\mathcal{N}}_{AD\to B E J_O})^{\otimes n})\ket{\theta_{A_{1,\text{typ}}^n A_{2,\text{typ}}^n A^n D^n}},
\end{multline}
with $A_{1,\text{typ}}^n \to R' G'_E G'_B \hat{G}'_{1, A} \hat{G}'_{2, A}$ and $A_{2,\text{typ}}^n \to R'' G''_B \hat{G}''_{A}$.

\subsubsection{Relay encoder and decoder}

We begin with the relay encoder. For each block $j-1$, there exists a family of partial isometries $\Lambda_{\tilde{\ubar{M}} \hat{\ubar{G}}_E^n \to {D}^n \ubar{G}_E}$ such that the following approximation holds:
\begin{multline}
    \ket{\theta_{\ubar{R}' \ubar{G}'_E \ubar{G}'_B \hat{\ubar{G}}'_{1,A} \hat{\ubar{G}}'_{2,A} \ubar{R}'' \ubar{G}''_B \hat{\ubar{G}}''_A \ubar{A}^n {D}^n}} \fqswEpsilon{4\sqrt{\varepsilon_2}}\\ \left[\Lambda_{\tilde{\ubar{M}}' \hat{\ubar{G}}_E^{'n} \to {D}^n \ubar{G}'_E}\otimes \Id \right]
    \bigl( \Phi_{\ubar{R}'\tilde{\ubar{M}}'} \otimes \Phi_{\ubar{R''}\ubar{M}''} \otimes \Phi_{\ubar{G}'_{1,A} \ubar{G}'_E}\\ \otimes\Phi_{\ubar{G}'_{2,A} \ubar{G}'_B}\otimes \Phi_{\ubar{G}''_A \ubar{G}''_B} \otimes\phi_{\ubar{B}^n \ubar{J}_O^n} \bigr),
\end{multline}
where $\theta_{D^n}\fqswEpsilon{\varepsilon_1} \sigma_D^{\otimes n}$, based on \App{sec:rel-state-split}. This is important for the next stage. 

As for the relay decoder, invoking the results of \App{sec:rel-state-split}, the partial isometry $\Gamma_{E^n G'_E \rightarrow \tilde{M}' \hat{G}_E^{'n}}$ satisfies
\begin{align}
    &\Phi_{R'\tilde{M}'} \otimes\Phi_{\hat{G}'_{1,A} \hat{G}_E^{'n}}\otimes \phi_{\hat{G}'_{2,A} G'_B R'' G''_B \hat{G}''_A  B^n J_O^n} \fqswEpsilon{4\sqrt{\varepsilon_2}}\nonumber\\ &\left[\Gamma_{E^n G'_E \to \tilde{M}' \hat{G}_E^{'n}}\otimes \Id \right] \ket{\phi_{R' G'_E G'_B \hat{G}'_{1, A} \hat{G}'_{2, A} R'' G''_B \hat{G}''_{A} B^n E^n J_O^n}},
\end{align}
where the quantum system $\tilde{M}'$ is the quantum ``message recovery" system at the relay decoder with $\rho_{R'\tilde{M}'}\approx\Phi_{R'M'}$.

\subsubsection{Destination Decoder}

Similarly to the above, according to \App{sec:receive-state-split}, for each block $j$, there exists a family of partial isometries $\Delta_{B^n G'_B G''_B \rightarrow \hat{M}' \hat{M}'' \hat{G}_B^{'n} \hat{G}_B^{''n}}$ such that:
\begin{multline}
    \Phi_{R' \hat{M}'}  \otimes \Phi_{R'' \hat{M}''} \otimes \Phi_{\hat{G}'_{2,A} \hat{G}_B^{'n}} \otimes \Phi_{\hat{G}''_A \hat{G}_B^{''n}}  \otimes\psi_{\hat{G}'_{1,A} G'_E E^n J_O^n} \fqswEpsilon{4\sqrt{\varepsilon_3}}\\ \left[\Delta_{B^n G'_B G''_B \to \hat{M}' \hat{M}'' \hat{G}_B^{'n} \hat{G}_B^{''n}}\otimes \Id \right]\cdot\\\ket{\omega_{R' G'_E G'_B \hat{G}'_{1, A} \hat{G}'_{2, A} R'' G''_B \hat{G}''_{A} B^n E^n J_O^n}},
\end{multline}
where the quantum systems $\hat{M}'$ and $\hat{M}''$ are the quantum ``message recovery" systems at the destination receiver, for the messages sent either directly or with the relay's help,  producing $\approx\Phi_{R'\hat{M}'}\otimes\Phi_{R''\hat{M}''}$ between Alice and Bob.
\qed}

{
\section{Proof of capacity bounds}\label{sec:proof-theorems-2-3-4}

In this section we prove Theorems \ref{thm:full-dec-thm-ent-no-ass}-
\ref{thm:full-dec-thm-ent-with-ass}, based on our result in Proposition~\ref{prop:part-dec-rates}.

\subsection{Proof of \Thm{thm:full-dec-thm-ent-no-ass} (Full decode-forward)}\label{sec:proof-full-dec-no-ass}

In the full decode-forward scheme, the auxiliary register $A_2$ is not needed. Hence, the initial state is $\ket{\sigma_{A A_1 D}}$. Taking the entanglement assistance rates to be $L'_{E},
\hat{L}'_{E}$, 
$L'_B$, $
\hat{L}'_B\rightarrow 0$, and combining the constraints in Eq. \eqref{eq:init-state-split-Q-ass}, \eqref{eq:relay-state-split-Q-ass}, and \eqref{eq:receive-state-relay-split-Q-ass}  yields the following bound:
\begin{equation}
    Q < \min \left\{
    I(A_1\rangle B)_\omega, I(A_1\rangle E)_\omega \right\} - \delta.
\end{equation}
The inequality in \Eq{eq:init-state-split-Q-ass} is inactive, since 
$
    H(A_1 | D)_\sigma
    =I(A_1\rangle A)_\sigma
$ by entropy duality, and $I(A_1\rangle A)_\sigma\geq I(A_1\rangle B)_\omega$ by the data processing inequality for the coherent information (see Theorem 11.9.3 in \cRef{ref:Wilde2017}). This concludes the proof of \Thm{thm:full-dec-thm-ent-no-ass}.

\subsection{Proof of \Thm{thm:part-dec-thm-ent-no-ass} (Partial decode-forward)}\label{sec:proof-part-dec-no-ass}

In the partial decode-forward scheme, the initial state is $\ket{\sigma_{A A_1 A_2 D}}$. Similarly to the proof of \Thm{thm:full-dec-thm-ent-no-ass}, we take the entanglement assistance rates to be $L'_{E}$, $
\hat{L}'_{E}
$, $L'_B$, 
$\hat{L}'_B$, 
$L''_B$,  $
\hat{L}''_B\rightarrow 0$. The constraints in Eq. \eqref{eq:init-state-split-Q-ass}, \eqref{eq:relay-state-split-Q-ass}, and \eqref{eq:receive-state-relay-split-Q-ass} yield
\begin{equation*}
    Q <  \min \bigl\{ I(A_1\rangle B)_\omega \,,\; I(A_1\rangle E)_\omega\bigr\}  + I(A_2\rangle B  A_1)_\omega - \delta.
\end{equation*}
Using the entropy chain rule, we may also write this as
\begin{equation}
    Q <  \min \bigl\{ I(A_1 A_2\rangle B)_\omega \,,\; I(A_1\rangle E)_\omega  + I(A_2\rangle B  A_1)_\omega \bigr\} - \delta.
\end{equation}
Similarly to the proof of \Thm{thm:full-dec-thm-ent-no-ass}, the inequality in \Eq{eq:init-state-split-Q-ass} is inactive by the data processing inequality. This concludes the proof for \Thm{thm:part-dec-thm-ent-no-ass}.

\subsection{Proof of \Thm{thm:full-dec-thm-ent-with-ass} (Entanglement assistance)}\label{sec:proof-part-dec-with-ass}

For the entanglement-assisted case of the full decode-forward strategy, we assume that only Bob has access to entanglement assistance. Therefore, we set the relay's entanglement rates to be $L'_{E}$, $
\hat{L}'_{E}\rightarrow 0$. Similarly to the proof of \Thm{thm:full-dec-thm-ent-no-ass}, the auxiliary register $A_2$ is not needed, and the initial state is $\ket{\sigma_{A A_1 D}}$. Isolating $Q$ in \Eq{eq:init-state-split-Q-ass} and \Eq{eq:receive-state-relay-split-Q-ass}, we achieve
\begin{equation}
    Q < \frac{1}{2}I(A_1 ; B)_\omega - \frac{1}{2}I(A_1 ; D)_\sigma - \delta.
\end{equation}
\qed}

\end{appendices}

{
 \balance
\bibliography{references}
}

\end{document}